\documentclass{article}

\usepackage{arxiv_style}

\usepackage{graphicx} % use graphic (ex, png etc..)
\usepackage{amsmath, amsfonts}
\usepackage{hyperref}       % hyperlinks
\usepackage{url}            % simple URL typesetting
\usepackage{nicefrac}       % compact symbols for 1/2, etc.
\usepackage{booktabs}       % professional-quality tables
\usepackage{microtype}      % microtypography

\title{Simplifying MBA Expression Using E-Graphs}
\author{
 Seoksu Lee \\
  Chungnam National University\\
  Daejeon\\
  Republic of Korea \\
  \texttt{troy.doubles@o.cnu.ac.kr} \\
   \And
 Hyeongchang Jeon \\
  Chungnam National University\\
  Daejeon\\
  Republic of Korea \\
  \texttt{jk365a@naver.com} \\
  \And
 Eun-Sun Cho \\
  Chungnam National University\\
  Daejeon\\
  Republic of Korea \\
  \texttt{eschough@cnu.ac.kr} \\
}

\begin{document}
\maketitle

\begin{abstract}
Code obfuscation involves the addition of meaningless code or the complication of existing code in order to make a program difficult to reverse engineer. In recent years, MBA (Mixed Boolean Arithmetic) obfuscation has been applied to virus and malware code to impede expert analysis. Among the various obfuscation techniques, Mixed Boolean Arithmetic (MBA) obfuscation is considered the most challenging to decipher using existing code deobfuscation techniques.
In this paper, we have attempted to simplify the MBA expression. We use an e-graph data structure to efficiently hold multiple expressions of the same semantics to systematically rewrite terms and find simpler expressions. The preliminary experimental result shows that our e-graph based MBA deobfuscation approach works faster with reasonable performance than other approaches do.
\end{abstract}


\section{Introduction}
In recent years, a number of program protection techniques have been emerged. One of these is program obfuscation, where the program is modified into a complicated form in order to secure the program. This method thwarts reverse engineering by third parties by making the program difficult to analyze. On the other hand, code obfuscation is also popular among malware distributors, who use it impede security analysts from analyzing such code.
Popular source code obfuscation tools include CodeVirtualizer, Themida, and VMprotect~\cite{CodeVirtualizer,Themida,VMProtect}, which perform code obfuscation by either inserting meaningless code into the code or transforming it into more complex code that has the same meaning as the existing code.
To analyze the obfuscated code, analysts have conducted research to find semantically equivalent simple code for the obfuscated code. Typically, solvers such as SMT-Solver and Z3-Solver~\cite{Z3solver} have been utilized to perform de-obfuscation by discovering less complex code that executes the same function semantically.
However, in response, malicious users have further advanced from conventional obfuscation techniques to effectively defeat solvers. In particular, they often use the MBA obfuscation technique~\cite{MBA_Blast}, which is not easily simplified by traditional solver-based de-obfuscation. This increases the cost of reverse engineering and analyzing malware. In this paper, we use an e-graph data structure~\cite{E_graph_1980, E_graph_2005}, which efficiently holds multiple expressions of the same semantics to systematically rewrite terms and find simpler expressions.

\section{Background}
\subsection{MBA (Mixed Boolean Arithmetic) expression}
MBA (Mixed Boolean Arithmetic) expression is a mixture of Boolean expressions and Arithmetic expressions.
MBA expressions are often generated by MBA obfuscation.
MBA obfuscation refers to expressing an expression given as an input in a more complex form by converting it into an expression mixed with Boolean operations that have the same meaning.

In the example below, a simple addition can be obfuscated by applying MBA obfuscation to replace the original expression with more complex MBA expression with a Boolean operation.
\[ a + b \Rightarrow (x \lor y) + y - (\neg x \land y) \]

Obfuscated MBA expression are classified into linear and polynomial expressions as defined below according to the form of the expression.

\textbf{Definition 1.}
    A linear MBA expression is defined \\
    \[\sum_{i \in I} a_{i}e_{i}(x_{1}, \ldots , x_{t})\] \\
    $a_{i}$ is a coefficient (integer constant) \\
    $e_{i}$ is a bitwise expression of variables $x_{1}, \ldots , x_{t}$ \\
    The bitwise operators include: inclusive or $(\lor)$, and $(\land)$, exclusive or $(\oplus)$, not $(\neg)$. $a_{i}e_{i}$ is called a term. \\

\textbf{Definition 2.}
    A polynomial MBA expression is defined as follows; \\
    \[ \sum_{i \in I} a_{i} (\prod_{j \in J} e_{i,j}(x_{1}, \ldots , x_{t})) \] \\
    $a_{i}$ is a coefficient (integer constant) \\
    $e_{i,j}$ is a bitwise expression of variables $x_{1}, \ldots , x_{t}$ \\
    $a_{i}e_{i}$ is called a term. \\

An example of polynomial MBA expression is \[ xy+3(x \land y)+4(x \land \neg y)(x \lor y-2) \] theses expressions take along time to analyze when performing the existing deobfuscation method and make it impossible to perform complete deobfuscation.
However, recent research has shown that in the case of linear MBA expressions, deobfuscation is possible to some extend.

\subsection{E-Graph}
Term Rewriting refers to changing a specific expression (term) to a simpler form by a rewriting rule. if there is a rewriting urle that matches the term, iterating the operation of substituting the exisiting term by that rule to find a simpler expression.
The resulting expression has the same meaning as the original complex expression but cant be expressed in a simpler form (normal form).

E-Graph~\cite{E_graph_1980,E_graph_2005} is a data structure proposed in 2005 as a graph data structure with the addition of equivalent class information (called e-class) with the same meaning.
This data structure can represent several terms with the same meaning at once, as shown in Figure~\ref{fig:egraph_explaination} below, and the simplest term can be quickly found in the graph represented.
It can allso be expressed simply when applying rewriting rules in the process of therm rewriting.

\begin{figure}[h]
    \centering
    \includegraphics[width=\textwidth]{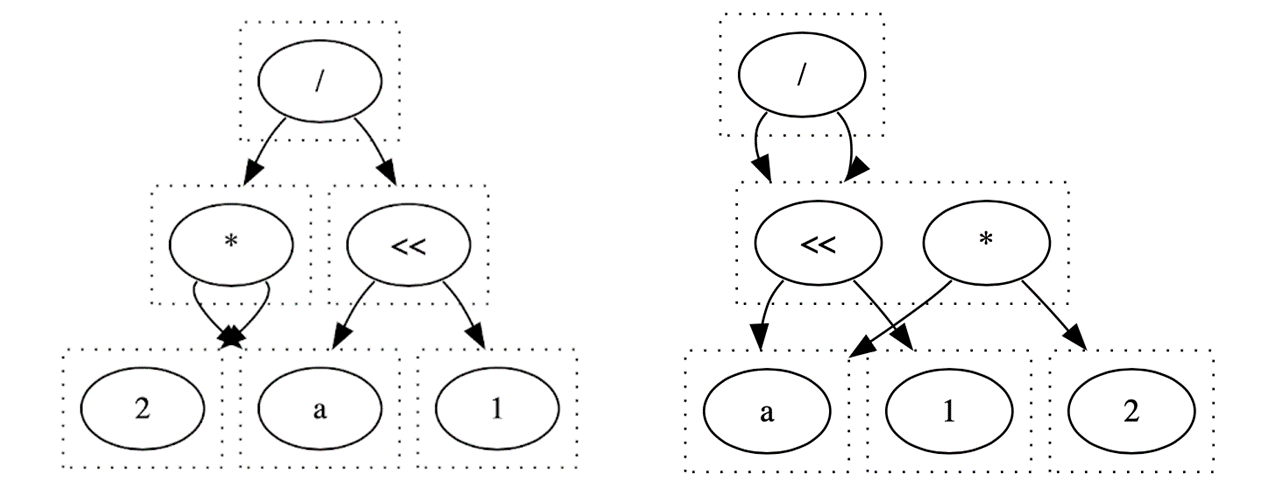}
    \caption{The figure on the left is an e-graph in the initial state representing the expression of ``(a * 2) / (a << 1)``, and after that, when the ``a << 1`` rewriting rule is applied, the figure on the right it is expressed in e-graph. The dotted line represents an e-class having the same semantic, and the solid circles represent nodes. This makes it easy to find the same e-class for parts that have the same semantics by the rewriting rule.}
    \label{fig:egraph_explaination}
\end{figure}

By using and E-graph, you can prevent incorrect ordering of rules during term rewriting, as well as missing infromation or adding incorrect semantics when perfoming substitutions.
Meanwhile existing MBA deobfuscation studies have been conducted using program synthesis, machine learning, and SMT Solver~\cite{Syntia, MBA_Solver, NeuReduce}.
These research results have their own merits and demerits depending on the MBA expression to be solved, but in general, there are drawbacks such as taking a long time or not being able to preserve the semantics.
In this paper, in order to fully preserve the meaning, we simplified the MBA expressino by applying rule-based term rewriting, and the simplified MBA expression can be used to effectively perform MBA deobfuscation by utilizing existing research.

\section{Implementation}
To simplify MBA expressions using E-graph, we utilized the Rust E-Graph library~\cite{egg}.
In the proess of using the library, the definition of MBA expressions was defined using Chapter 2, and the simplification process was implemented based on the direction of reducing the size of abstract syntax tree comprising the expression.
In addition, the following preprocessing tasks were applied to the MBA expressions used as input.

\begin{itemize}
  \item In order to distinguish between the (-) operation meaning substraction and logical negation ($\sim$) operation, we separate the operator.
  \item Converted the ($\sim$) negation operator to a (\& 0xffff) operation.
\end{itemize}

The rules used for term rewriting were implemented by utilizing basic rules used in mathmatics (commutative law, distribution law, Boolean operation, set.) and MBA deobfuscation papers.
If too many rules are applied, term rewriting in the reverse direction is applied in the process of finding the optimal value(term) during the simplification process, resulting in an increase in the search space or simplifying to one value with a completely different semantic from the previous one, so we used the basic rules.

\section{Experiment}
The results of simplification on the MBA expression dataset with the implemented simplifier are shown below Table~\ref{table:emba_experiment}.

\begin{table*}[ht]
    \centering
    \caption{MBA Expression Simpliciation Experiment Results}
    \resizebox{\textwidth}{!}{%
    \begin{tabular}{ccccccc}
    \toprule[1.5pt]
    \textbf{Dataset}       & \textbf{\# of Total expression} & \textbf{\# of Sucess} & \textbf{\# of Failure} & \textbf{Success Rate} & \textbf{Simplifcation Ratio} & \textbf{Time (s)} \\ \bottomrule[1.5pt]
    Tigress                & 323                             & 267                   & 56                     & 82.66\%               & 69\%                         & 3.98              \\
    Qsynth Custom EA       & 501                             & 493                   & 8                      & 98.40\%               & 65.67\%                      & 72.79             \\
    MBA\_Solver (Linear)   & 1008                            & 818                   & 190                    & 81.15\%               & 93.26\%                      & 41.13             \\
    MBA\_Solver (Non-poly) & 1003                            & 949                   & 54                     & 94.61\%               & 93.26\%                      & 239.04            \\
    MBA\_Solver (Poly)     & 1008                            & 587                   & 421                    & 58.23\%               & 94.91\%                      & 27.15             \\ \bottomrule[1.5pt]
    \end{tabular}%
    }
    \label{table:emba_experiment}
\end{table*}
  
The MBA expression datasets used in the experiment are Tigress, Qsynth customEA, and MBA Solver data.
For MBA Solver data the data contains linear, non-polynomial, and polynomial MBA expressions.
As shown in ~\ref{emba_experiment}, we were able to simplify most of the expressions, and the time was within 0.1 seconds on average for one MBA expression.
Among them, we can see that MBA Solver polynomial expresions have a higher number of failures compared to other obfuscation datasets because there are fewer rules associated with them.
We can also see that the expressions used in the Tigress and Qsynth customEA datasets are simplified to 70\%, which is about 30\% less than original MBA expressions.
However, we can also see that the MBA Solver dataset does not simplify the simplified expressions significantly compared to the origin expressions due to the large number of constant values.
To compare the performance, a comparative experiment was conducted usgin SSPAM and GAMBA from previous studies.
In the case of SSPAM, most of the MBA obfsucation expressions could not be simplified, and it was confirmed that both performance and time decreased as the expressions became longer, such as exceeding 5 minutes for single expression.
In the case of GAMBA, which was studied most recently, we can see that the Qsynth customEA data failed to simplify the same 8 expressions and successfully simplified all other exrpessions for the dataset we used.
However, we, can see where the spped up was relatively slow, with 576 seconds for the Qsynth customEA data and 126 seconds for the Tigress data.
  
\section{Conclusion}
In this paper, we proposed a method to simplify the MBA expression with the term rewriting technique using E-Graph.
It can be a good approach to solve the time problem that exists in the existing MBA deobfuscation tool.
In the future, we aim to develop an effective MBA deobfuscation tool based on E-Graphs by adding effective rules to simplify Polynomial MBA expressions that are not successful with this tool and applying constant folding to further simplify the expression.

\section{ACKNOLEDGEMENTS}
This work was supported by Institue for Information \& communications Technology Planning \& Evaluation (IITP) grant funded by the Korea government (MSIT) (No.2022-0-01200, Convergence security core talent traning business (Chungnam National University)
We would like to express our deep appreciation to Prof. Woosuk Lee. at Hanyang Univ. for his valuable suggestions.
  
\bibliographystyle{unsrt}
\bibliography{references}

\end{document}